\documentclass[%
reprint,
showpacs,preprintnumbers,
 amsmath,amssymb,
 aps,
prb,
]{revtex4-1}
\usepackage{graphicx}
\usepackage{dcolumn}
\usepackage{bm}
\usepackage{hyperref}
\usepackage{amsmath}
\usepackage{amsfonts}
\usepackage{mathtools,bm}
\usepackage[amssymb]{SIunits}
\usepackage{color}
\usepackage{mathrsfs}
\usepackage{soul}
\usepackage{fancyhdr}
\usepackage{lineno}
\usepackage{leftidx}
\usepackage[yyyymmdd,hhmmss]{datetime}
\newcommand*\patchAmsMathEnvironmentForLineno[1]{%
  \expandafter\let\csname old#1\expandafter\endcsname\csname #1\endcsname
  \expandafter\let\csname oldend#1\expandafter\endcsname\csname end#1\endcsname
  \renewenvironment{#1}%
     {\linenomath\csname old#1\endcsname}%
     {\csname oldend#1\endcsname\endlinenomath}}%
\newcommand*\patchBothAmsMathEnvironmentsForLineno[1]{%
  \patchAmsMathEnvironmentForLineno{#1}%
  \patchAmsMathEnvironmentForLineno{#1*}}%
\AtBeginDocument{%
\patchBothAmsMathEnvironmentsForLineno{equation}%
\patchBothAmsMathEnvironmentsForLineno{align}%
\patchBothAmsMathEnvironmentsForLineno{flalign}%
\patchBothAmsMathEnvironmentsForLineno{alignat}%
\patchBothAmsMathEnvironmentsForLineno{gather}%
\patchBothAmsMathEnvironmentsForLineno{multline}%
}

\begin{document}

\title{Polaritonic linear dynamic in Keldysh formalism}
\author{A.~Rahmani}
\email{rahmani@stu.yazd.ac.ir}
\affiliation{Physics Department, Yazd University, P.O. Box 89195-741, Yazd, Iran}
\author{M.~A.~Sadeghzadeh}
\affiliation{Physics Department, Yazd University, P.O. Box 89195-741, Yazd, Iran}
\date{\today}

\begin{abstract}
  We study the dynamic of polaritons in the Keldysh functional
  formalism. Dissipation is considered through the coupling of the
  exciton and photon fields to two independent photonic and excitonic
  baths. As such, this theory allows to describe more intricate decay
  mechanisms that depend dynamically on the state of the system, such
  as a direct upper-polariton lifetime, that is motivated from
  experiments.  We show that the dynamical equations in the Keldysh
  framework otherwise follow the same Josephson--like equations of
  motions than the standard master equation approach, that is however
  limited to simple decay channels.  We also discuss the stability of
  the dynamic and reconsider the criterion of strong coupling in the
  presence of upper polariton decay.
\end{abstract}

\maketitle
\section{Introduction}

Many-body phenomena in the strong coupling regime of light--matter
interactions, in particular Bose-Einstein condensation, have attracted
considerable attention in recent
years\cite{deng10a,kavokin_book11a}. The photon, one of the intrinsic
components of the polariton, has an inevitable interaction with the
environment, which leads to its decay and often affects the dynamic in
a non-negligible way. A well--studied example of non--equilibrium
quantum transition takes place in microcavity~\cite{imamoglu96a}. A
semiconductor microcavity with embedded low dimensional structures
provides a unique laboratory to study a variety of quantum
phases. This advantage finds its existence due to a composite boson:
the \emph{exciton-polariton}~\cite{hopfield58a,weisbuch92a}, a quantum
superposition of light and matter with not-only fermionic but also a
photonic component.  The polaritonic side of the light--matter
coupling has stimulated research in both fundamental and applied
fields. With regard to the applicability, polaritons promise devices
with remarkable upgrades as compared to their semiconductor
counterparts, from which polariton
lasers~\cite{imamoglu96a,christopoulos07a,daskalakis13a,azzini11a} and
polariton transistors~\cite{amo10a,ballarini13a,anton14a,gao15a} are
the most obvious examples. Concerning fundamental aspects, polaritons
cover immense areas in physics, including Bose-Einstein Condensation
(BEC)~\cite{kasprzak06a,deng07a,balili07a,lai07a},
superfluidity~\cite{carusotto04a,wouters10b}, spin Hall
effect~\cite{kavokin05b}, superconductivity~\cite{laussy10a} and
Josephson--effects
effects~\cite{lagoudakis10a,abbarchi13a,voronova15b,Rahmani16a}.

\begin{figure}[t]
  \centering
  \includegraphics[width=.7\linewidth]{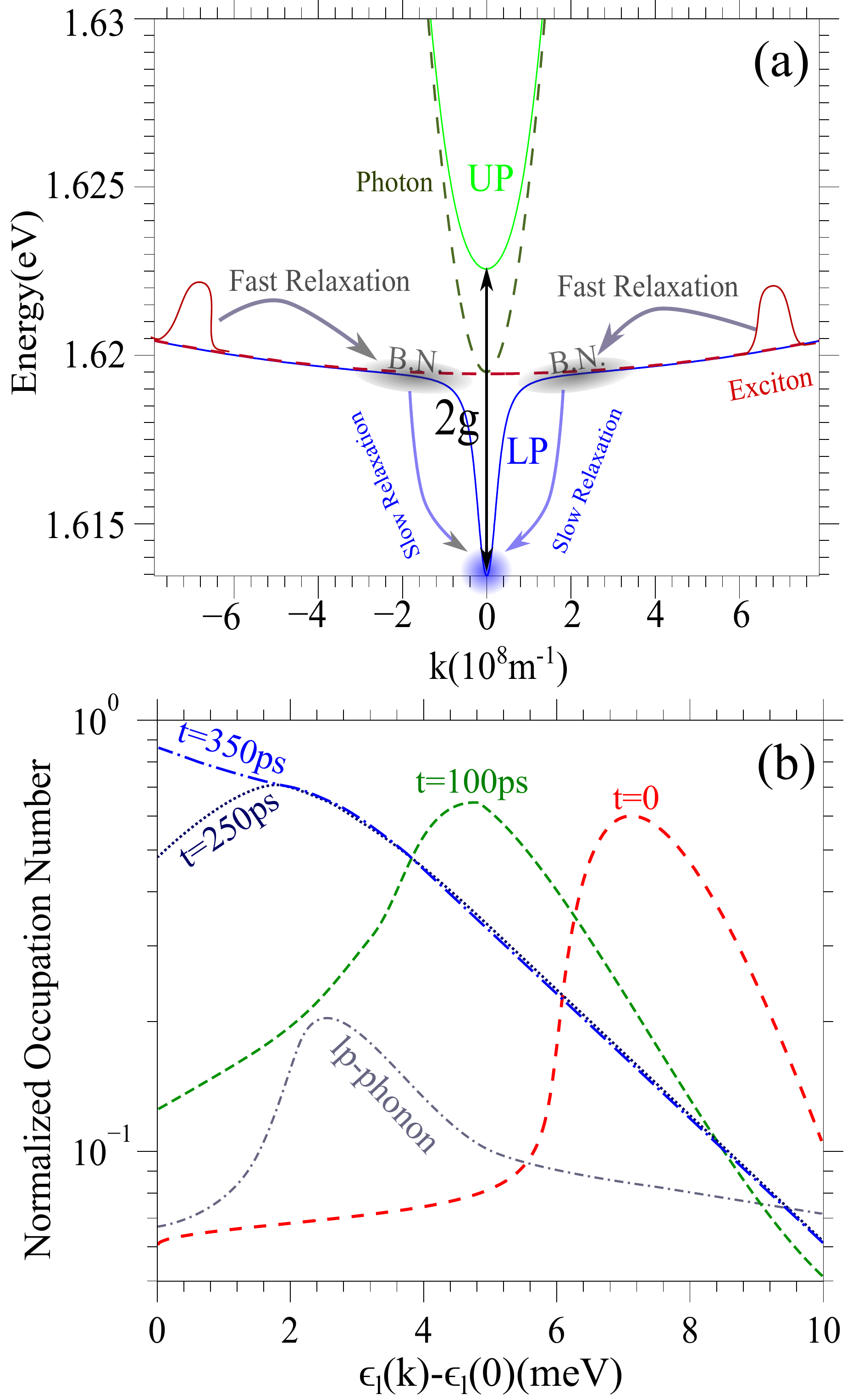}
  \caption{(a)~Dispersion of polariton when the detuning at~$k=0$~is
    zero ($\epsilon_a(0)=\epsilon_b(0)$). Strong coupling between
    photon (in dash-dark green) and exciton (in dashed red) results
    the lower polariton (denoted by Lp and in blue) and the upper
    polariton (denoted by Up and in light green). The energy splitting
    at~$k=0$~is~$2g$. In the regime of weak pumping, polaritons are
    accumulated at the bottleneck region (B.N.). By increasing the
    pumping strength, the B.N. is relaxed and the condensate
    forms. The relaxation of polaritons due to interactions with other
    polariton or phonons is fast in exciton--like region, but is slow
    in photon--like region. (b)~Distribution function of
    polariton. The initial non--degenerate distribution~(red dashed
    line)~evolves to a degenerate distribution (blue dashed--dot
    line), with both polariton--polariton and polariton--phonon
    interactions. Considering the phonon--polariton processes only,
    the distribution remains non--degenerate (gray dashed--dot
    line). Using parameters for simulation are taken form
    Ref.~\cite{doan05a}.}
  \label{fig:marjul7120100CEST2015115sd8}
\end{figure}

The simplest description for the kinetic of a polariton gas is
provided by semiclasical Boltzmann equations~\cite{snoke10b}. This
approach has been used widely by many
authors~\cite{tassone99a,porras02a,malpuech02a,doan05a,hartwell10a,maragkou10a}. Due
to the fast photon leakage from the microcavity, polaritons have a
short lifetime, which keeps the polaritonic system in nonequilibrium
regime. Therefore the system should be pumped to compensate the losses
of polaritons. The rate equation for condensation kinetics of
polaritons is:
\begin{align}
  \partial_tn_\mathbf{k}=&p_\mathbf{k}-\frac{n_\mathbf{k}}{\tau_\mathbf{k}}+\frac{\partial n_\mathbf{k}}{\partial t}\vert_{lp-lp}+\frac{\partial n_\mathbf{k}}{\partial t}\vert_{lp-ph}\,,\label{eq:3}
\end{align}  
where~$p_\mathbf{k}$~is the pumping term,
and~$(\tau_\mathbf{k})^{-1}$~describes the particle decay rate.  The
two next terms in Eq.~(\ref{eq:3}) account for polariton--polariton
and polariton--phonon scattering rates, respectively. We refer to
Appendix~\ref{app:Boltzmanneq} for detailed calculations of scattering
rates. To illustrate the formation of the condensate in polaritonic
system, we present a typical result of numerical simulation of the
Boltzmann equation in
Fig~\ref{fig:marjul7120100CEST2015115sd8}(b). Initially, polaritons
are introduced incoherently in exciton--like region of the lower
polariton dispersion~(Fig.~\ref{fig:marjul7120100CEST2015115sd8}(a)),
and then relax quickly, except near the exciton--photon resonance. In
this region, the polariton density of states is reduced and the
photonic contribution to the polariton is increased, which results in
polariton accumulations in the bottleneck
region~\cite{tassone97a}. This effect is shown as the peak in the
curve in Fig.~\ref{fig:marjul7120100CEST2015115sd8}(b).  To make the
population degenerate, that is to overcome the bottleneck effect, one
needs to take into account the action of both polariton--polariton and
polariton--phonon processes, as the only polariton--phonon scattering
mechanism arises pilled up population in non--zero state. With both
scattering processes, then Bose stimulation effectively amasses
polaritons in the ground state.

While it is a simple, though extremely time consuming, simulation,
Boltzmann equations exclude the quantum aspect of the dynamic, to wit,
it does not consider the effect of coherence. One then needs to
upgrade the formalism to include both quantum and non--equilibrium
aspects of the dynamic. A powerful and widely used method that allows
such an exact treatment is provided by the Keldysh functional integral
approach~\cite{kamenev_book11a,sieberer16a}. This method has been
applied to study the driven open system including polaritons in
microcavities~\cite{szymanska06a,szyma07a,proukakis13a,Dunnett16a,pavlovic13a},
glassy and supperradiant phase of ultracold atoms in optical
cavity~\cite{buchholda13,Torre13a}, photon condensations in dye filled
optical cavity~\cite{leeuwa13}, etc. It also allows to explore the
Bose--Hubbard model with time-dependent hopping~\cite{kennetta11}, and
non--equilibrium Bosonic Josephson oscillations~\cite{trujilloa09}, among others.

In this paper, a quantum field theory for polariton internal degrees
of freedoms in dissipative regime using Keldysh functional method is
developed. The need for such an approach in describing the polariton
relaxation is motivated mainly by the polariton specificity of
providing two types of lifetimes~\cite{dominici14a}: one for the bare
states (exciton and photon), the other for the dressed states (upper
polariton)~\cite{skolnick00a}. In the latter case, excitons and
photons are removed in a correlated way from the system. This effect
has attracted attention recently in particular as it can lead to
interesting applications proper to polaritons~\cite{colas15a}.  While
this polariton lifetime can be described in a linear regime from
standard master equation approaches, more general scenarios involving,
e.g., interactions, get out of reach of this description as the nature
of the dressed state changes dynamically with the state of the
system. A description of the dissipation therefore needs being made at
a more fundamental level than through the standard Lindblad
phenomenological form.

While photon and exciton are considered as independent quantum field
interacting through the Rabi energy, to model the decay, we assume two
independent excitonic and photonic baths, which are present in most
light--matter coupled systems.  In this text, we restrict our analysis
to the linear (Rabi) regime (no interaction). The interaction
certainly has important effects, such as
self--trapping~\cite{raghavan99a}, and the optical parametric
oscillator regime~\cite{Dunnett16a}. However, it can be found that
even in the non-interacting regime of the dynamic, some aspects of
nonlinearity emerge from the exciton-photon
coupling~\cite{rubo03a,voronova15a,Rahmani16a}. Therefore, we first
attack the problem of polariton dynamics in the Keldysh formalism in
the simpler case of non-interacting particles, as a basis for more
elaborate and involved studies. At such, we derive the mean--field
equations of motions in the photon-exciton
basis~\cite{hampa15,elistratova16,voronova15a}. In particular, we show
that the polaritonic internal dynamic satisfies the Josephson
criterion of coherent flow.

This paper is organised as follows. In Sec.~\ref{sec:two} we present
the polaritonic Hamiltonian and how to turn it into a dissipative
system. This includes the Hamiltonian for coupling both bare and
dressed fields baths. The Keldysh technique is introduced in
Sec.~\ref{sec:three}, where the mean-field solutions and fluctuation
actions are also presented. In Sec.~\ref{sec:four}, we represent the
internal dynamic on the Paria sphere~\cite{Rahmani16a} (dynamically
renormalized Bloch sphere), and discuss on the stability of the
solution. Conclusions are presented in Sec.~\ref{sec:five}.
\section{Polariton Hamiltonian}
\label{sec:two}

The strong coupling between photon and exciton fields in a
semiconductor microcavity results in a quasiparticle with very
peculiar properties: the polariton. Denoting the photon and exciton
field operators by~$a_\mathbf{k}$~and~$b_\mathbf{k}$~respectively,
then the Hamiltonian describing the internal coupling between the two
fields is given by:
\begin{subequations}
\begin{align}
\label{eq:Hc}
H_c=&H_0+H_{Rabi}\,,\\
\label{eq:H0}
H_0=&\sum_\mathbf{k}(\epsilon_a(k)a^\dagger_\mathbf{k} a_\mathbf{k}+\epsilon_b(k)b^\dagger_\mathbf{k} b_\mathbf{k})\,,\\
\label{eq:HR}
H_{Rabi}=&\sum_\mathbf{k}g(a^\dagger_\mathbf{k}b_\mathbf{k}+b^\dagger_\mathbf{k}a_\mathbf{k})\,,
\end{align}
\end{subequations}
where~$\epsilon_a$~and~$\epsilon_b$~are the cavity photon and quantum
well exciton dispersion given respectively by:
\begin{subequations}
\begin{align}\label{eq:photondispersion}
\epsilon_a(k)=&\frac{\hbar c}{\sqrt{\varepsilon}}\sqrt{k^2_\perp+k^2}\,,\\
\epsilon_b(k)=&\epsilon^{ex}(0)+\frac{\hbar^2k^2}{2m_{ex}}\,,
\end{align}
\end{subequations}
with~$\epsilon^{ex}(0)=2m_{ex}e^4/\varepsilon^2\hbar^2$~as the 2D
exciton binding energy. In Eq.~(\ref{eq:HR}), $g$~shows the strength of
coupling between photon and exciton fields and in the regime of strong
coupling, it is referred to as the Rabi energy.

Diagonalising the Hamiltonian in Eq.~(\ref{eq:Hc}) leads to the new
bosonic dressed modes: the lower~($L_\mathbf{k}$)~and
upper~($U_\mathbf{k}$)~polariton. Then the~$H_c$~takes the diagonal
form:
\begin{align}\label{eq:indiagonalform}
H_c=\sum_\mathbf{k}\epsilon_lL_\mathbf{k}^\dagger L_\mathbf{k}+\epsilon_uU_\mathbf{k}^\dagger U_\mathbf{k}\,,
\end{align}
with~$\epsilon_{\underset{l}{u}}=\frac12(\epsilon_a+\epsilon_b\pm\sqrt{(\epsilon_a-\epsilon_b)^2+(2g)^2}$.
Note that for zero detuning ($\epsilon_a(0)=\epsilon_b(0)$), the
splitting the two polariton branches is~$2g$. The dispersion in zero
detuning is shown in Fig.~(\ref{fig:marjul7120100CEST2015115sd8}).

Such a transformation from bare states (photon and exciton) to dressed
states (upper and lower polariton) is done through the operator
relation:
\begin{subequations}
\begin{align}
L_\mathbf{k}=&A(k)a_\mathbf{k}+B(k)b_\mathbf{k}\,,\\
U_\mathbf{k}=&B(k)a_\mathbf{k}-A(k)b_\mathbf{k}\,,
\end{align}
\end{subequations}
where~$A(k)$~and~$B(k)$~are the so called Hopfield
coefficients~\cite{hopfield58a}, given by~\cite{ciuti03a,laussy04b}:
\begin{subequations}
\begin{align}
B(k)=\frac{1}{\sqrt{1+(\frac{g}{\epsilon_{lp}-\epsilon_c})^2}}\,,\\
A(k)=-\frac{1}{\sqrt{1+(\frac{\epsilon_{lp}-\epsilon_c}{g})^2}}\,,
\end{align}
\end{subequations}
Due to photon leakage from the microcavity, the polariton has a short
lifetime. To consider the dynamic in dissipative regime, one should
begin from a microscopic view of the mechanism underling dissipation,
namely to model the \emph{environmental interaction} by coupling the
system to a bath. Here the Hamiltonian for the undamped system is
given in Eq.~(\ref{eq:H0}), while the baths are modeled as a
collection of harmonic oscillators:
\begin{align}\label{Hbath}
H_{bath}=&\sum_{\mathbf{p}}\omega_\mathbf{p}^{ph}r^\dagger_\mathbf{p}r_\mathbf{p}+\sum_{\mathbf{p}}\omega_\mathbf{p}^{ex}c^\dagger_\mathbf{p}c_\mathbf{p}\,,
\end{align}
with~$\omega_\mathbf{p}^{ph(ex)}$~as the dispersion of the photonic
(excitonic) bath, and corresponding creation and annihilation
operators~$r^\dagger_\mathbf{p}(c^\dagger_\mathbf{p})$~and~$r_\mathbf{p}(c_\mathbf{p})$,
respectively. It is assumed that each bath is in thermal equilibrium
and unaffected by the behavior of the system. The bath--system
interaction can be described through:
\begin{align}
\label{eq:Hdey}
H_{Dec}=\sum_{\mathbf{k,p}}[(F_\mathbf{k}^\mathbf{p}a^\dagger+R_\mathbf{k}^\mathbf{p}b^\dagger)r+(S_\mathbf{k}^\mathbf{p}a^\dagger+G_\mathbf{k}^\mathbf{p}b^\dagger)c]+\mathrm{H.c.}\,,
\end{align}
where~$\mathrm{H.c.}$~stands for Hermitian conjugate. Parameters in
Eq.~(\ref{eq:Hdey})~are related to the coupling between polaritonic
system and baths which are defined as:
\begin{align}
F_\mathbf{k}^\mathbf{p}&\equiv \Gamma_{\mathbf{k},ph}^\mathbf{p}+B\Gamma_{\mathbf{k},u}^\mathbf{p}\,,\hspace{5mm}R_\mathbf{k}^\mathbf{p}\equiv -A\Gamma_{\mathbf{k},u}^\mathbf{p}\nonumber\,,\\
G_\mathbf{k}^\mathbf{p}&\equiv \Gamma_{\mathbf{k},ex}^\mathbf{p}-A\Gamma_{\mathbf{k},u}^\mathbf{p}\,,\hspace{5mm}S_\mathbf{k}^\mathbf{p}\equiv B\Gamma_{\mathbf{k},u}^\mathbf{p}\nonumber\,,
\end{align}
where~$\Gamma_{\mathbf{k},ph(ex)}^\mathbf{p}$~shows the coupling
strength of the photonic (excitonic) component of the polarion to the
photonic (excitonic) bath. The polariton can also decay through its
upper branch, which is modeled via direct coupling to the both baths
with coupling strength of~$\Gamma_{\mathbf{k},u}$. Deriving all the
needed Hamiltonian we find for our final Hamiltonian:
\begin{align}
  \label{eq:Hamiltonian+in+Linear+regime}
  H=&H_c+H_{Dec}+H_{bath}\,.
\end{align}
\section{Functional representation of polaritons}
\label{sec:three}

In this section, we present the functional approach to the internal
dynamic of polaritons. Any equilibrium many--body theory involves
adiabatic switching on of \emph{interaction} at a distant past
($t=-\infty$), and off at a distant future ($t=\infty$). The state of
the system at these two reference times is the ground state of the
non--interacting system. Then any correlation function in the
interaction representation can be averaged with respect to a known
ground state of the non--interacting Hamiltonian.

The postulate of independence of the reference states from the details
of switching on and off the interaction breaks in non--equilibrium
condition, as the system evolves to an unpredictable state. However,
one needs to know the final state. It was Schwinger~\cite{scwa60}'s
suggestion that the final state to be exactly is the same as that of
the initial time. Then the theory can evolve along a two--branch
closed time contour with a forward and backward direction.

The central quantity in the functional integral method is the
partition function of the system that can be written as a Gaussian
integral over the bosonic fields of~$\phi,\bar{\phi}$:
\begin{align}\label{eq:partionfun}
Z=&N~\int D[\bar{\phi},\phi]~e^{iS[\bar{\phi},\phi]}\,,
\end{align}
where~$N$~is the normalization constant and~$S$~is the action, which
carries the dynamical information.  In the Keldysh formalism, the
bosonic field~$\phi$~is split into two
components~$\phi^+$~and~$\phi_-$, which reside on the forward and
backward part of the time contour. Then the field are rotated to the
Keldysh basis defined as:
\begin{align}\label{eq:keldyshrotation}
\phi_{\underset{q}{cl}}=&\frac{1}{\sqrt{2}}(\phi_+\pm\varphi_-)\,,
\end{align}
where the~$+(-)$~sign stands for~$cl(q)$. Here~$cl(q)$~stands for
classical (quantum) component of the field.

Corresponding to the terms in the
Hamiltonian~(\ref{eq:Hamiltonian+in+Linear+regime}), the actions take
the following components in Keldysh space:
\begin{align}\label{eq:S0}
S_0=&\Delta_\mathbf{k}^t[\Psi_\mathbf{k}^\dagger(i\partial_t-\epsilon_a)\sigma_1^K\Psi_\mathbf{k}+\Phi_\mathbf{k}^\dagger(i\partial_t-\epsilon_b)\sigma_1^K\Phi_\mathbf{k}]\,,
\end{align}
\begin{align}\label{eq:SR}
S_{Rabi}=&-\Delta_\mathbf{k}^{t}g~[\Phi_\mathbf{k}^\dagger\sigma_1^K\Psi_\mathbf{k}+\Psi_\mathbf{k}^\dagger\sigma_1^K\Phi_\mathbf{k}]\,,
\end{align}
\begin{align}\label{eq:SDec}
S_{Dec}^{r}=&-\Delta_\mathbf{k,p}^{t}~[(F_\mathbf{k}^\mathbf{p}\Psi^\dagger_\mathbf{k}+R_\mathbf{k}^\mathbf{p}\Phi^\dagger_\mathbf{k})\sigma^K_1X_{R,\mathbf{p}}+\mathrm{h.c.}]\,,
\end{align}
\begin{align}\label{eq:SDecwx}
S_{Dec}^{c}=&-\Delta_\mathbf{k,p}^{t}~[(S_\mathbf{k}^\mathbf{p}\Psi^\dagger_\mathbf{k}+G_\mathbf{k}^\mathbf{p}\Phi^\dagger_\mathbf{k})\sigma^K_1X_{C,\mathbf{p}}+\mathrm{h.c.}]\,,
\end{align}
\begin{align}\label{eq:Sbath}
S_{bath}=&\Delta_\mathbf{p}^{t}~\sum_{j=R,C}X_{j,\mathbf{p}}^\dagger(i\partial_t-\omega_\mathbf{p}^j)\sigma_1^KX_{j,\mathbf{p}}\nonumber\\=&\Delta_\mathbf{p}^{t}~\sum_{j=R,C}X_{j,\mathbf{p}}^\dagger C_{j}^{-1}X_{j,\mathbf{p}}\,, 
\end{align}
where we use~$\Delta_\mathbf{k}^{t}$~as an abbreviation
for~$\sum_\mathbf{k}\int_{-\infty}^\infty~dt$, and
superscript~$r$~and~$c$~refer to excitonic and photonic baths,
respectively. Other notations are summarised in Table~\ref{tab:one}.
\begin{table}[b]
\centering\caption{Fields and matrices in Keldysh space}
\label{tab:one}
\begin{tabular}{@{}|c|c|c|c|@{}}\hline
\multicolumn{4}{@{}|c|@{}}{\rule[-0.125cm]{0mm}{0.5cm}%
\mbox{Fields in Keldysh space}}\\
\hline
\mbox{Photonic}&
\mbox{Ecxitonic} & \mbox{Photonic Bath} & \mbox{Excitonic Bath}\\ \hline
$\Psi_\mathbf{k}$
   &$\Phi_\mathbf{k}$&$X_{c,\mathbf{k}}$&$X_{r,\mathbf{k}}$\\
\hline$\left( \begin{array}{c}
\psi_{cl}\\\psi_{q} 
\end{array}\right)_\mathbf{k}$&$\left( \begin{array}{c}
\varphi_{cl}\\\varphi_{q} 
\end{array}\right)_\mathbf{k}$&$\left( \begin{array}{c}
x_{cl}\\x_{q} 
\end{array}\right)_\mathbf{k}$&$\left( \begin{array}{c}
y_{cl}\\y_{q} 
\end{array}\right)_\mathbf{k}$
    \\ \hline
\multicolumn{4}{@{}|c|@{}}{\rule[-0.125cm]{0mm}{0.5cm}%
\mbox{Pauli matrices in Keldysh space}}\\
\hline
\mbox{$\sigma_K^0$}&\mbox{$\sigma_K^1$}&
\mbox{$\sigma_K^2$} & \mbox{$\sigma_K^3$}\\ \hline
$\left( \begin{array}{cc}
1&0\\0&1
\end{array}\right)$
   &$\left( \begin{array}{cc}
   0&1\\1&0
   \end{array}\right)$&$\left( \begin{array}{cc}
   0&-i\\i&0
   \end{array}\right)$&$\left( \begin{array}{cc}
      1&0\\0&-1
      \end{array}\right)$\\\hline
\end{tabular}
\end{table}

The bath action in Eq.~(\ref{eq:Sbath}) is in the standard form in
Keldysh space: it contains a quadratic form of the fields with a
matrix which is the inverse of a
correlator~$C_j$~with~$j=\left\lbrace r,c\right\rbrace $. One can
show that:
\begin{align}
C_j\equiv&\left( \begin{array}{cc}
C^K_i&C^R_j\\
C^A_j&0
\end{array}\right)_{t,t'}\nonumber\\
=&\left( \begin{array}{cc}
-if_je^{-i\omega_p^j(t-t')}&-i\Theta(t-t')e^{-i\omega_p^j(t-t')}\\
i\Theta(t-t')e^{-i\omega_p^j(t-t')}&0
\end{array}\right)\,,
\end{align}
where~$\Theta(t)$~is the Heaviside step function and~$f_j$~is the
distribution function of the bath.

As the bath coordinates appear in a quadratic form, they can be
integrated out to reduce the degree of freedom to photon and exciton
fields only. We follow the procedure described in
Refs.~\cite{kamenev_book11a,szyma07a}. Employing the properties of
Gaussian integration, the decay bath eliminating leads to two
effective actions:
\begin{subequations}
\begin{align}
\label{eq:SDecwithoutbath1}
S_{Dec}^{r}=&\Delta_\mathbf{k,p}^{t,t'}~(F_\mathbf{k}^\mathbf{p}\Psi^\dagger_\mathbf{k}+R_\mathbf{k}^\mathbf{p}\Phi^\dagger_\mathbf{k})_{t}~\mathcal{C}^{-1}_r(t-t')~(F_\mathbf{k}^\mathbf{p}\Psi_\mathbf{k}+R_\mathbf{k}^\mathbf{p}\Phi_\mathbf{k})_{t'}\,,\\
\label{eq:SDecwithoutbath11}
S_{Dec}^{c}=&\Delta_\mathbf{k,p}^{t,t'}~(S_\mathbf{k}^\mathbf{p}\Psi^\dagger_\mathbf{k}+G_\mathbf{k}^\mathbf{p}\Phi^\dagger_\mathbf{k})_{t}~\mathcal{C}^{-1}_c(t-t')~(S_\mathbf{k}^\mathbf{p}\Psi_\mathbf{k}+G_\mathbf{k}^\mathbf{p}\Phi_\mathbf{k})_{t'}\,,
\end{align}
\end{subequations}
where~$\mathcal{C}^{-1}_{j}(t-t')=-\sigma_1^K~[C_{j}(t-t')]~\sigma_1^K$~and~$j=\left\lbrace
  r,c\right\rbrace$.
Straightforward matrix multiplication shows that
the~$\mathcal{C}^{-1}$~correlator has the causality structure, given
by:
\begin{align}
\mathcal{C}^{-1}_j(t-t')\equiv&\left( \begin{array}{cc}
0&\mathcal{C}^A_j\\
\mathcal{C}^R_j&\mathcal{C}^K_j
\end{array}\right)_{t,t'}\,.
\end{align}

To proceed further, we make some simplifying assumptions about the
baths. Firstly, we assume that all modes of the systems are coupled to
their baths with the same strength, i.e,
$\Gamma_j(p)\equiv\Gamma_{\mathbf{k},j}^\mathbf{p}$. Besides, it is
assumed that the bath is in the Markovian limit, where the density of
state for the baths and the coupling between the system and baths are
constant.  In the following, we restrict our analysis to these
assumptions. More details including non--Markovian cases are presented
in Appendix~\ref{app:freqencydependence}.

Denoting the decay action
as~$S_{Dec}=\sum_{i=r,c}S_{bath}^i+S_{Dec}^i$, one gets the components
of the~$S_{D}$~as (see Appendix~\ref{app:freqencydependence}):
\begin{align}\label{eq:sDecay1}
S_{Dec}^1=&\sum_\mathbf{k}\int~dt\{ \gamma_1(k)\Psi^\dagger_\mathbf{k}\sigma_2^K\Psi_\mathbf{k}\\\nonumber&+2i(\gamma_1(k))\int dt'\bar{\psi}_q(t)(f_c+f_d)_{t-t'}\psi_q(t')\}\,,
\end{align}
\begin{align}\label{eq:sDecay2}
S_{Dec}^2=&\sum_\mathbf{k}\int~dt\{ \gamma_2(k)\Phi^\dagger_\mathbf{k}\sigma_2^K\Phi_\mathbf{k}\\\nonumber&+2i(\gamma_2(k))\int dt'\bar{\varphi}_q(t)(f_c+f_d)_{t-t'}\varphi_q(t')\}\,,
\end{align}
\begin{align}\label{eq:sDecay3}
S_{Dec}^3=&\sum_\mathbf{k}\int~dt\{ \gamma_3(k)\Psi^\dagger_\mathbf{k}\sigma_2^K\Phi_\mathbf{k}\\\nonumber&+2i(\gamma_3(k))\int dt'\bar{\psi}_q(t)(f_c+f_d)_{t-t'}\varphi_q(t')\}\,,
\end{align}
\begin{align}\label{eq:sDecay4}
S_{Dec}^4=&\sum_\mathbf{k}\int~dt\{ \gamma_3(k)\Phi^\dagger_\mathbf{k}\sigma_2^K\Psi_\mathbf{k}\\\nonumber&+2i(\gamma_3(k))\int dt'\bar{\varphi}_q(t)(f_c+f_d)_{t-t'}\psi_q(t')\}\,,
\end{align}
where we
define~$\gamma_1=B^2\gamma_u^2+(\gamma_a+B\gamma_u)^2$,~$\gamma_2=A^2\gamma_u^2+(\gamma_b-A\gamma_u)^2$,~$\gamma_3=\gamma_u(B(\gamma_b-A\gamma_u)-A(\gamma_a+B\gamma_u))$,
and~$\sigma^{K}_2$~is given in Table~\ref{tab:one}. One notes that
with decay for the upper polariton, the decay action has the
weightings of excitonic and photonic Hopfield coefficients; moreover,
even without bare field couplings~$\gamma_a$~and~$\gamma_b$, the decay
in the upper branch is enough to remove the photon and exciton fields
both independently and in a correlated way.

\subsection{Mean field solutions}

Having integrated out the bath degree of freedom, the action appears
in its final form as: $S=S_0+S_R+S_{Dec}$. We can then obtain the
equations of motions from the saddle point condition on the action:
\begin{align}
\frac{\partial S}{\partial\bar{\psi}_{i=q,cl}}=0\,,\hspace{5mm}\frac{\partial S}{\partial\bar{\varphi}_{i=q,cl}}=0\,.
\end{align}
which yields:
\begin{subequations}
\begin{align}
\label{eq:equationmotion1}
\frac{\partial S}{\partial\bar{\psi}_q}=&(i\partial_t-\epsilon_a+i\gamma_1)\psi_{cl}-(g-i\gamma_3)\varphi_{cl}\nonumber\\&+2i\gamma_1\int~dt'(f_c+f_d)(t-t')\psi_q(t')\nonumber\\&+2i\gamma_3\int~dt'(f_c+f_d)(t-t')\varphi_q(t')\,,\\
\label{eq:equationmotion2}
\frac{\partial S}{\partial\bar{\varphi}_q}=&(i\partial_t-\epsilon_b+i\gamma_2)\varphi_{cl}-(g-i\gamma_3)\psi_{cl}\nonumber\\&+2i\gamma_2\int~dt'(f_c+f_d)(t-t')\varphi_q(t')\nonumber\\&+2i\gamma_3\int~dt'(f_c+f_d)(t-t')\psi_q(t')\,,\\
\label{eq:equationmotion3}
\frac{\partial S}{\partial\bar{\psi}_{cl}}=&(i\partial_t-\epsilon_a-i\gamma_1)\psi_{q}-(g+i\gamma_3)\varphi_{q}\,,\\
\label{eq:equationmotion4}
\frac{\partial S}{\partial\bar{\varphi}_{cl}}=&(i\partial_t-\epsilon_b-i\gamma_2)\varphi_{q}-(g+i\gamma_3)\psi_{q}\,.
\end{align}
\end{subequations}
One notices that
Eqs.(\ref{eq:equationmotion3}-\ref{eq:equationmotion4}) are satisfied
by~$\varphi_q=\bar{\varphi}_q=0$~and~$\psi_q=\bar{\psi}_q=0$,
irrespective of what the classical
components,~$\varphi_0\equiv\varphi_{cl}$~and~$\varphi_0\equiv\varphi_{cl}$,
are. Then, under these conditions,
Eqs.~(\ref{eq:equationmotion1}-\ref{eq:equationmotion2}) lead to:
\begin{align}\label{eq:coupledequations1}
\partial_t\psi_{0}=&(-i\epsilon_a-\gamma_1)\psi_{0}+(-ig-\gamma_3)\varphi_{0}\,,
\end{align}
\begin{align}\label{eq:coupledequations2}
\partial_t\varphi_{0}=&(-i\epsilon_b-\gamma_2)\varphi_{0}+(-ig-\gamma_3)\psi_{0}\,.
\end{align}
These equations provide an extreme limit for the action~$S$ and
describe the system of two coupled-equations of motions in the mean
field analysis.

\subsection{Fluctuations of the action}

By separating the fluctuations from the mean field:
\begin{align}
\psi=\psi_0+\delta\psi_{cl}\,,\hspace{5mm}\psi_q=\delta\psi_q\,,\\
\varphi=\varphi_0+\delta\varphi_{cl}\,,\hspace{5mm}\varphi_q=\delta\varphi_q\,,
\end{align}
for the actions defined in
Eqs.~(\ref{eq:S0})--(\ref{eq:SR})~and~(\ref{eq:sDecay1})--(\ref{eq:sDecay4}),
then employing the Fourier transform, the fluctuating action takes the
form:
\renewcommand\arraystretch{1.3}
\begin{align}\label{eq:DeltaS}
\Delta S=\int~d\omega\sum_\mathbf{k}\Delta\mathcal{T}_\mathbf{k}^\dagger\left( \begin{array}{c|c}
0&[\mathcal{F}^{-1}]^A\\\hline
[\mathcal{F}^{-1}]^R&[\mathcal{F}^{-1}]^K
\end{array}\right) \Delta\mathcal{T}_\mathbf{k}\,,
\end{align}
where the superscripts~$A$,~$R$, and~$K$~stand for the advanced,
retarded and Keldysh components of the inverse Green function,
respectively. The fluctuation vector has the form of:
\begin{align}
\Delta\mathcal{T}_\mathbf{k}^\dagger\equiv(CL,Q)^T
\end{align}
with
\begin{align}
CL\equiv\left( \begin{array}{c}
\delta\bar{\psi}_{cl}(\omega)\\
\delta\psi_{cl}(-\omega)\\
\delta\bar{\varphi}_{cl}(\omega)\\
\delta\varphi_{cl}(-\omega)
\end{array}\right)_\mathbf{k}\quad\text{and}\quad Q\equiv\left( \begin{array}{c}
\delta\bar{\psi}_{q}(\omega)\\
\delta\psi_{q}(-\omega)\\
\delta\bar{\varphi}_{q}(\omega)\\
\delta\varphi_{q}(-\omega)
\end{array}\right)_\mathbf{k}\,.
\end{align}
Further, the off--diagonal matrix elements in Eq.~(\ref{eq:DeltaS})
have the following relations~\cite{kamenev_book11a}:
\begin{align}\label{eq;relationamonggreenF1}
[\mathcal{F}^{-1}]^{R(A)}~\mathcal{F}^{R(A)}=1\,,\\
\mathcal{F}^A=(\mathcal{F}^R)^\dagger\,,
\end{align}
while for the diagonal element one finds:
\begin{align}\label{eq;relationamonggreenF2}
\mathcal{F}^K=\mathcal{F}^RF-F\mathcal{F}^A
\end{align}
where ~$F$~is referred to as the distribution function of the
system~\cite{kamenev_book11a}. Having found the Green functions of the
dynamic, one can decide about the stability of the solution by
studying the retarded Green function, namely by
solving~$\det([\mathcal{F}^{-1}]^R(\omega_r))=0$, where~$\omega_r$~is
the pole of the retarded Green
function. If~$\mathrm{Im}[\omega_r]<0$, then the proposed solution is
stable.
\begin{figure*}[t]
  \centering
  \includegraphics[width=\linewidth]{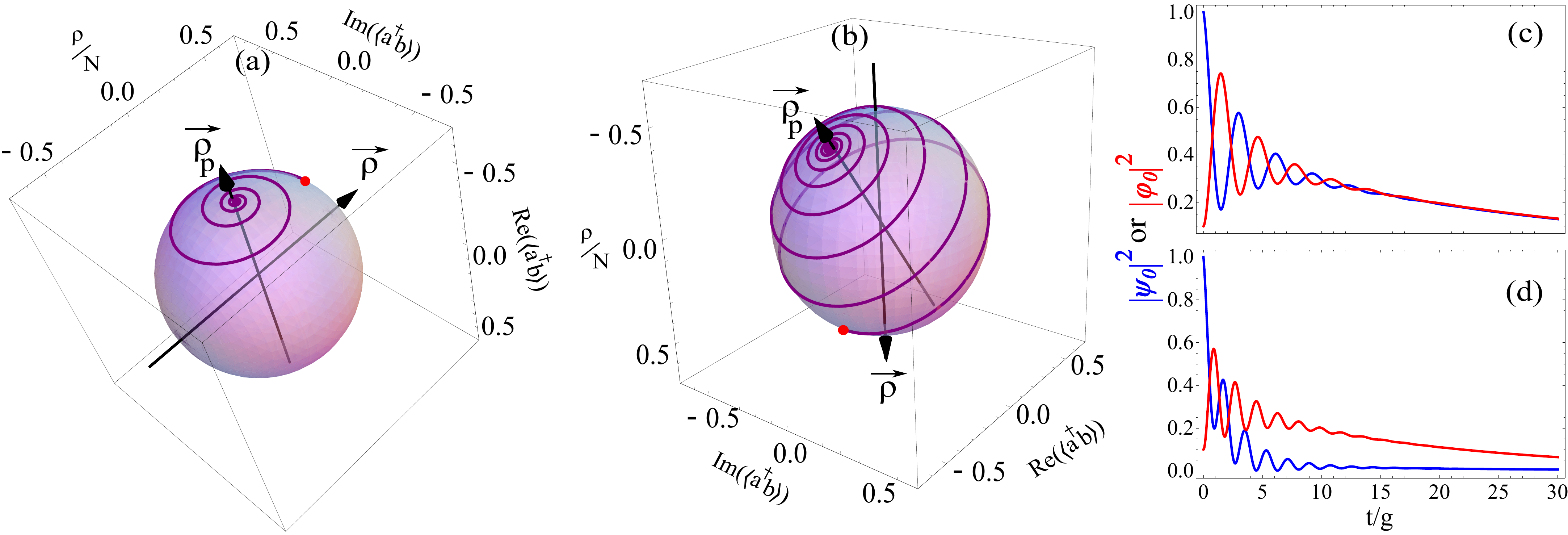}
  \caption{(a) The internal dynamic of polariton in a photon--like
    point on the Paria sphere, when detuning between exciton and
    photon fields is zero. The~$\rho_p$~direction shows the
    orientation of the upper--lower polariton states, while
    the~$\rho$~shows the direction of exciton--photon states. The red
    point shows the initial point. (b)~The same as (a) but for an
    exciton--like point. (c)~and~(d)~shows the population dynamic in
    photon and exciton like point respectively. Using parameters
    are:~$\gamma_a=0.2g$,~$\gamma_b=0.02g$,~$\gamma_u=0.3g$.}
  \label{fig:marjul7120100CEST20151158}
\end{figure*}
\section{Discussion}\label{sec:four}

To analyse the equations of motions, we start from
Eqs.~(\ref{eq:coupledequations1}) and~(\ref{eq:coupledequations2}),
and restrict calculations to two points in reciprocal space: the space
center ($k=0$), that is a photon--like point, and an exciton--like
point at~$k\sim\sqrt{2m_{ex}\epsilon_{ex}^{0}}$. In the absence of
exciton and photon detuning at~$k=0$, the coupling fields are at
resonance, and the exciton and photon have the same weight of~$1/2$ in
the polariton. However, bare states at~$k\neq0$ are positively
detuned, which provides an intrinsic detuning in the internal dynamic
of polaritons.

Introducing~$\rho_\mathbf{k}\equiv\frac12(|\psi_0|^2-|\varphi_0|^2)$ as
the population imbalance
and~$N_\mathbf{k}\equiv(|\psi_0|^2+|\varphi_0|^2)$ as the total field
population in state~$\mathbf{k}$, one can show that the
Eqs.~(\ref{eq:coupledequations1}) and~(\ref{eq:coupledequations2}) are
in the form of Josephson equations, that is:
\begin{subequations}
\label{eq:inrhoandsigma}
\begin{align}
\partial_t(\rho/N)_\mathbf{k}=&-\sqrt{1-4(\rho/N)_\mathbf{k}^2}(g\sin(\sigma_\mathbf{k})-\gamma_3\cos(\sigma_\mathbf{k})(\rho/N)_\mathbf{k})\nonumber\\&-(\gamma_1+\gamma_2)(1-4(\rho/N)_\mathbf{k}^2)\,,\\
\partial_t\sigma_\mathbf{k}=&-\delta_\mathbf{k}\nonumber\\&+\frac{1}{\sqrt{1-4(\rho/N)_\mathbf{k}^2}}(4g\cos(\sigma_\mathbf{k})(\rho/N)_\mathbf{k}+\gamma_3\sin(\sigma_\mathbf{k}))\,,
\end{align}
\end{subequations}
where~$\delta_\mathbf{k}=\epsilon_a-\epsilon_b$ stands for detuning
between the states, and~$\sigma_\mathbf{k}=\arg[\psi_0^\ast\varphi_0]$
is the relative phase between bare states. With such a representation
of the dynamic, each polaritonic state in~$\mathbf{k}$~has an
intrinsic internal Josephson--like dynamic, when the relative phase
drives the population difference. Recently, the same equations of
motions were reported by one of the authors, but for an effective
two-level system and in a different
formalism~\cite{Rahmani16a,Rahmani16wolfram}. One notices that the
phase difference between the two coupled fields, with a definite phase
in each field, is crucial to drive the internal dynamic. This is the
case when two or more condensates are coupled, through for example
Josephson junctions. However, here we do not take any assumption about
the condensate phase, and this is left to the initial condition to
define a clear phase for each field; then as long as the polaritonic
system is initially prepared by a source of specific phase, the
internal dynamic follows the Josephson dynamic, and the coherence
oscillates between the fields.

An example of the dynamic is shown in
Fig.~(\ref{fig:marjul7120100CEST20151158}). Here we adopt the Paria
sphere~\cite{Rahmani16wolfram} to observe the dynamic in a three
dimensional representation. Two directions are indicated on the
sphere: the~$\vec\rho$~direction, that shows the direction for
exciton--photon states, and~$\vec\rho_p$, which shows the orientation
of lower--upper polariton states.  Starting from an initial point (the
red point in Fig.~(1)), the dynamic goes toward a fixed point (when
the sphere is kept normalized). In the zero detuning case,
(Fig~(\ref{fig:marjul7120100CEST20151158}--a)), the laboratory basis
is orthogonal to the dressed state basis,
with~$\vec\rho\perp\vec\rho_p$, and the relative phase remains in an
oscillatory mode. Going toward an exciton--like point in
Fig~(\ref{fig:marjul7120100CEST20151158}--b), the two directions are
not orthogonal, that shows the final state has a more exciton
weight. At the same time, one can see a switching from the running
mode to the oscillatory mode in relative phase, which is mediated by
decay. We also show the dynamic of bare state populations in
photon--like (Fig.~(\ref{fig:marjul7120100CEST20151158}--c)) and
exciton--like(Fig.~(\ref{fig:marjul7120100CEST20151158}--d)) points of
the indirect space. We set the initial conditions to have more
population in the photon field. At zero detuning, both fields are
oscillating in the same trend, as the decay affects the dynamic
equivalently; however, by increasing the detuning, decay affects the
field (in this example the photon field) that has the more
population. In other words, by increasing the detuning, bare states
become decoupled and each field loses its coupling to other fields
while the coupling to the bath is yet active.
\begin{figure}[b]
  \centering
  \includegraphics[width=.7\linewidth]{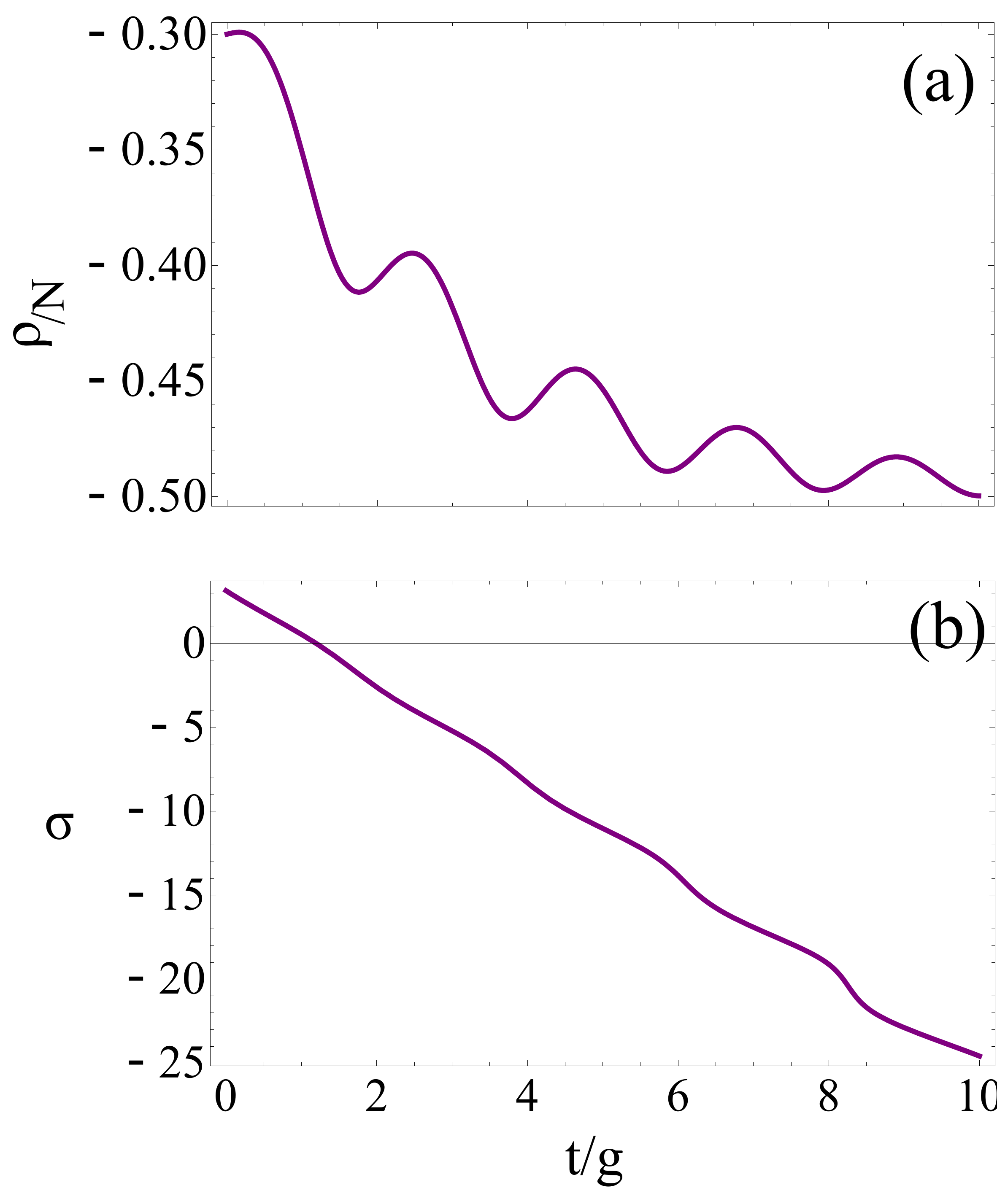}
  \caption{(a) damped oscillations in population imbalance mediated by
    decay in the upper polariton. Corresponding relative phase is
    shown in (b).  Parameters used
    are:~$\delta=3g,~\gamma_a=\gamma_b=.2g,~\gamma_u=.6g,~\rho(0)=-0.3N,~\sigma(0)=\pi$.}
  \label{fig:marjul7120100CEST2015115sd8s}
\end{figure}

The equations of motions in their Josephson representation bring a new
variant for the dynamics of polariton. Careful inspection in
Eqs.~(\ref{eq:inrhoandsigma}) shows that the relative phase is driving
the population in two ways: one is the well--studied Josephson dynamic
with the term proportional to~$g\sin(\sigma_\mathbf{k})$ in the
equations form Ref.~\cite{raghavan99a}. The other comes from the term
proportional to~$\gamma_3\cos(\sigma_\mathbf{k})(\rho/N)_\mathbf{k}$,
which existence is related to the upper polariton decay, and is
specific to this phenomenon.  Such a peculiar aspect of the dynamic
holds even for the case of disconnected fields which are correlatively
coupled to a bath. A particular example of the internal dynamic
mediated only by polariton decay is shown in
Fig.~(\ref{fig:marjul7120100CEST2015115sd8s}). As the relative phase
is in the running mode, the population imbalance exhibit damped
oscillations toward a fixed point.

In any dynamical system, the stability of the solution in the steady
state is the most important property. In normalized coordinates, the
fixed points of the dynamic have a finite values, even in the
dissipative regime~\cite{Rahmani16a}. For a given fixed point, the
stability condition is determined through the inverse retarded Green
function in Eq.~(\ref{eq:DeltaS}), which reads:
\begin{widetext}
\begin{align}\label{InverseRetardedGreen}
[\mathcal{F}^{-1}]^R(\omega,\mathbf{k})=\frac{1}{2}\left( \begin{array}{cccc}
\omega-\epsilon_a+i\gamma_1&0&-g+i\gamma_3&0\\
0&-\omega-\epsilon_a-i\gamma_1&0&-g-i\gamma_3\\
-g+i\gamma_3&0&\omega-\epsilon_b+i\gamma_2&0\\
0&-g-i\gamma_3&0&-\omega-\epsilon_b-i\gamma_2
\end{array}\right)\,.
\end{align}
\end{widetext}
One notices how coupling between photon and exciton fields are
reflected as the coupling between quantum and classical parts of the
fields in Keldysh space, which makes the retarded matrix
non--diagonal. More interestingly is the appearance of the
term~$\gamma_3$, which has the same weight in the dynamic as the
coupling-constant~$g$. This is the direct consequence of the decay in
the upper polariton branch, as it removes bare fields in a correlated
fashion.

Solving~$[\det(\mathcal{F}^{-1}]^R(\omega_r)=0$, one finds the energies
of the dressed states in the dissipative regime as:
\begin{align}
\omega_r^{\pm}=&\frac12[\epsilon_a+\epsilon_b-i(\gamma_1+\gamma_2)\nonumber\\&\pm\sqrt{(\delta_\mathbf{k}-2ig-\gamma_+)(\delta_\mathbf{k}+2ig-\gamma_-)}]\,,
\end{align}
where~$\delta_\mathbf{k}=\epsilon_a-\epsilon_b$ is the detuning, and~$\gamma_\pm=\pm2\gamma_3+i(\gamma_1-\gamma_2)$. The version with no dissipation has the familiar form~\cite{elistratova16}:
\begin{align}
\omega_r^{\pm}=&\frac12(\epsilon_a+\epsilon_b\pm\sqrt{\delta_\mathbf{k}+4g^2})\,,
\end{align}
that is the result from a pure Hamiltonian picture (see the notes
after Eq.~\ref{eq:indiagonalform}). For zero detuning, one gets:
\begin{align}
\omega_r^{\pm}=&\frac12[\epsilon_a+\epsilon_b-i(\gamma_1+\gamma_2)\nonumber\\&\pm\sqrt{(-2ig-\gamma_+)(+2ig-\gamma_-)}]\,.
\end{align}
For~$\gamma_3=0$, the term under the square root can go from real to
imaginary, namely, it happens when~$g<(\gamma_1-\gamma_2)/2$, which is
considered as the criterion for strong to weak coupling
transition~\cite{laussy08a,kavokin_book11a}. However,
for~$\gamma_3\neq0$, we see clearly that the term under the square
root remains imaginary, which breaks the criterion of strong-to-weak
coupling transition at zero detuning. The imaginary parts
of~$\omega^\pm_r$ determine the stability of the
solutions. Straightforward calculations lead to:
\begin{align}
  \mathrm{Im}[\omega_r^\pm]=-(\gamma_1+\gamma_2)-\sqrt{\frac{-\mathbb{IM}+\sqrt{\mathbb{IM}^2+\mathbb{RE}^2}}{2}}\,,
\end{align}
where~$\mathbb{IM}=-2\delta_\mathbf{k}(\gamma_1-\gamma_2)-8g\gamma_3$~and~$\mathbb{RE}=\delta_\mathbf{k}^2-(\gamma_1-\gamma_2)^2+4(g^2-\gamma_3^2)$. Clearly,
it can be seen that the imaginary part of the~$\omega_r^\pm$, for
given parameters of the system, always remains negative, which results
in stable solutions. One direct consequence of such stability is the
resistance of the system against phase transitions, which is the case
in presence of interactions and pumping. Clearly, the combination of
decay and interaction makes the dynamics richer and their full effects
will be discussed in future works.

\section{Conclusion}
\label{sec:five}

In conclusion, we study the internal dynamic of polariton in the
Keldysh functional approach, when fields are removed from both bare
and dressed states. In the linear Rabi regime, the coupled equations
of motions are local in reciprocal space, and the intrinsic detuning
between bare states works as an intrinsic potential affecting the
dynamic. It is shown also that the equations of motions are in the
form of Josephson equations, but that the upper-polariton lifetime
(correlated decay of the dressed state) brings a peculiar feature in
the dynamics, namely, it mediates an internal dynamic between the bare
states. This would happen even if the bare states would be decoupled
(although then the origin for their correlated decay would be less
clear on physical grounds). Considering the retarded Green functions,
we show that the dynamic in the Rabi regime is stable, and the
criterion of strong coupling is fragile in presence of an upper
polariton decay.

\begin{acknowledgments}
  We thanks F.P. Laussy for having suggested some aspects of this
  problem and for discussions.
\end{acknowledgments}

\appendix
\section{}\label{app:Boltzmanneq}

Here we describe two important scattering mechanisms in the polariton
kinetic. Main equations for numerical calculation are presented.

\subsection{polariton--polariton scattering}

Suppose the occupation number of state~$\mathbf{k}$~is given
by~$n_\mathbf{k}$, then the time variation of the occupation number
due to polariton--polariton interaction reads:
\begin{align}\label{eq:99}
\frac{\partial n_\mathbf{k}}{\partial t}\vert_{lp-lp}=&-n_\mathbf{k}\sum_{\mathbf{k'},\mathbf{q}}W^{lp-lp}_{\mathbf{k}'\mathbf{k}}(n_{\mathbf{k}'}+1)(n_{\mathbf{q}}+1)n_{\mathbf{k'}+\mathbf{q}-\mathbf{k}}\nonumber\\&+(n_\mathbf{k}+1)\sum_{\mathbf{k}',\mathbf{q}}n_{\mathbf{k}'}n_{\mathbf{q}}(n_{\mathbf{k}'+\mathbf{q-k}}+1)W^{lp-lp}_{\mathbf{k}\mathbf{k}'}\,,
\end{align}
where~$W^{lp-lp}_{\mathbf{k}\mathbf{k}'}$~is the polariton--polariton
scattering rate for transition of polariton form
state~$\mathbf{k}$~to~$\mathbf{k'}$. Using the Fermi's golden rule,
the scattering rate reads:
\begin{align}
W_{\mathbf{k,k'}}^{lp-lp}=\frac{2\pi}{\hbar} B_{eff}|M(|\mathbf{k-k'}|)|^2\delta(E_k+E_{k_s}-E_q-E_{k'})\,,\label{eq:10}
\end{align}
with~$B_{eff}=B(k)^2B(k')^2B(q)^2B(k_s)^2$~as the effective excitonic weight, and~$\mathbf{k}_s=\mathbf{k'+q-k}$. The exciton--exciton matrix element,~$M$, has been studied by Ciuti \emph{et al.}~\cite{ciuti98a} and recently by Sun \emph{et al.}~\cite{arXiv_sun}. Here we use the estimation provided by Tassone and Yamamoto~\cite{tassone99a} as~$M\approx 6a_B^2\epsilon^{ex}/A_s$, where~$a_B$~is the two dimensional Bohr radios of the exciton, and~$A_s$~is the area of the sample. Replacing the sum by integral (thermodynamic limit) and employing the properties of delta function one gets:
\begin{align}
\frac{\partial n_\mathbf{k}}{\partial t}|_{lp-lp}=&\frac{A_s^2}{8\pi^3\hbar}\int k' dk' q dq d\theta' B_{eff}M^2(\frac{\partial \epsilon_l}{\partial k_s^2})^{-1}\nonumber\\&\times[\frac{n_kn_{k_s}(1+n_q)(1+n_{k'})}{((c_{min}-\cos(\theta'))(\cos(\theta')-c_{max}))^{1/2}}\nonumber\\&+ \frac{n_{k'}n_q(1+n_{k_s})(1+n_{k})}{((c_{min}-\cos(\theta'))(\cos(\theta')-c_{max}))^{1/2}}]\,,\label{eq:final}
\end{align}
where~$c_{min}(c_{max})$~is the lower (upper) limit of integrations over~$\theta'$, and is of the form:
\begin{eqnarray}
c_{min(max)}=\frac{k^2+k'^2-(k_s\mp q)^2}{2kk'}\,.
\end{eqnarray}
\subsection{polariton--phonon scattering}

Polariton can scatter from one state to other states through emission or absorption of phonons. The time variation of the occupation number caused by polariton--phonon scattering reads:
\begin{align}
\frac{\partial n_\mathbf{k}}{\partial t}\vert_{lp-ph}=I_{in}^{abs}+I_{in}^{em}+I_{out}^{abs}+I_{out}^{em}\,,\label{eq:knonzero}
\end{align}
where the portion of emission and absorption of phonon in polariton scattering is shown by superscript~$em$~and~$abs$~respectively. Here we describe, for example,~$I_{in}^{em}$~in details. This term describes polariton scattering rate form~$\mathbf{k'}$~to~$\mathbf{k}$~while a phonon is absorbed. Other terms in Eq.~(\ref{eq:knonzero}) can be drived straightforwardly. Utilizing the Fermi's golden rule, one finds
\begin{equation}
I_{in}^{em}=(n_\mathbf{k}+1)\frac{2\pi}{\hbar}\sum_{\mathbf{k}',q_z}W^2n_\mathbf{k'}n_\mathbf{Q}~\delta(E_{k'}-E_k-E_Q)\,,\label{eq:Iinem}
\end{equation}
where we call~$\mathbf{Q}=(\mathbf{q=k-k'},q_z)$~the phonon wavevector, and~$W$~stands for transition probability. We restrict our analysis to longitudinal--acoustic phonon, for which the polariton--phonon interaction is provided by the deformation--potential coupling with electron and hole~$D_e$~and~$D_h$, correspondingly. Taking~$\mathbf{r}_{i=e,h}=(\pmb{\rho}_i,z_i)$~and~$\pmb{\rho}=\pmb{\rho}_e-\pmb{\rho}_h$~, the exciton wavefunction in state~$\mathbf{k}$~is given by~$\langle\mathbf{r}_e,\mathbf{r}_h|\mathbf{k}\rangle=\frac{1}{\sqrt{A_s}}\exp(i\mathbf{k}\cdot\mathbf{R})f(\mathbf{\rho})U_e(z_e)U_h(z_h)$. Then the transition probability reads 
\begin{equation}
W=B(k)B(k')\sqrt{\frac{\hbar Q}{2\rho_duV}}\mathbb{D}(|\mathbf{k-k'}|)\mathbb{A}(q_z)\delta_{\mathbf{q,k-k'}}\,,\label{eq:W}
\end{equation}
where~$\mathbb{A}=\int dz_ee^{iq_zz_e}U^2_e(z_e)$~and~$\mathbb{D}=D_eG(\beta_hq)+D_hG(-\beta_eq)$. The form factor~$G$~is defined~$G(x)=\int f^2e^{i\mathbf{x}\cdot\pmb{\rho}}d\pmb{\rho}$. Replacing the sum with integral in Eq.~(\ref{eq:Iinem}) we finally have
\begin{align}
I_{in}^{em}=\frac{(n_\mathbf{k}+1) V}{\hbar^4u^3(2\pi)^2}\int k'dk'd\theta\frac{W^2n_\mathbf{k'}n_\mathbf{Q}(\epsilon_l(k)-\epsilon_l(k'))^2}{\sqrt{(\frac{\epsilon_l(k)-\epsilon_l(k')}{\hbar u})^2-|\mathbf{k-k'}|^2}}\,.\label{eq:final1}
\end{align}
\section{}\label{app:freqencydependence}
To integrate over bath fields we take the vantage Gaussian integral~\cite{kamenev_book11a}. In the following we present the results for photonic bath. Calculation including excitonic bath can be done straightforwardly by replacing~$F^\mathbf{p}_\mathbf{k}$,~$F^\mathbf{p}_\mathbf{k}$~and~$\rho_r$~with~$S^\mathbf{p}_\mathbf{k}$,~$G^\mathbf{p}_\mathbf{k}$~and~$\rho_c$, respectively. Following the procedure described in Refs.~\cite{kamenev_book11a,szyma07a} one has (for photonic bath):
\begin{widetext}
\begin{subequations}
\begin{align}
S^r_{Dec}=&I_1+I_2+I_3\,,\\
I_1=&\Delta_{\mathbf{k,p}}^{t,t'}\left[ (F_\mathbf{k}^\mathbf{p})^2(\bar{\psi}_{cl}(t)C_r^A(t-t')\psi_q(t')+\bar{\psi}_{q}(t)C_r^R(t-t')\psi_{cl}(t')+\bar{\psi}_{q}(t)C_r^K(t-t')\psi_q(t'))\right] \,,\\
I_2=&\Delta_{\mathbf{k,p}}^{t,t'}[ F_\mathbf{k}^\mathbf{p}R_\mathbf{k}^\mathbf{p}(\bar{\psi}_{cl}(t)C_r^A(t-t')\varphi_q(t')+\bar{\varphi}_{cl}(t)C_r^A(t-t')\psi_q(t')+\bar{\psi}_{q}(t)C_r^R(t-t')\varphi_{cl}(t')\nonumber\\&+\bar{\varphi}_{q}(t)C_r^R(t-t')\psi_{cl}(t')+\bar{\psi}_{q}(t)C_r^K(t-t')\varphi_{q}(t')+\bar{\varphi}_{q}(t)C_r^K(t-t')\psi_{q}(t'))]\\
I_3=&\Delta_{\mathbf{k,p}}^{t,t'}[(R_\mathbf{k}^\mathbf{p})^2(\bar{\varphi}_{cl}(t)C_r^A(t-t')\varphi_q(t)+\bar{\varphi}_{q}(t)C_r^R(t-t')\varphi_{cl}(t)+\bar{\varphi}_{q}(t)C_r^K(t-t')\varphi_q(t))]\,.
\end{align}
\end{subequations}
Replacing the sum over~$\mathbf{p}$~with an integral $\sum_{\mathbf{p}}\rightarrow\int~d\xi~\rho_r(\xi)$, with~$\rho_r(\xi)$~as the bath density of states, one finds (after Fourier transformation)
\begin{subequations}
\begin{align}
S^r_{Dec}=&\int~d\omega[ \bar{\psi}_{cl}(\omega)(\Sigma^+_{ph}+B^2\Sigma^+_{u}-2B\Sigma^+_{ph-u})\psi_q(-\omega)+\bar{\psi}_{cl}(\omega)(A\Sigma^+_{ph-u}-AB\Sigma^+_{u})\varphi_{q}(-\omega)\nonumber\\&+\bar{\varphi}_{cl}(\omega)(A\Sigma^+_{ph-u}-AB\Sigma^+_{u})\psi_{q}(-\omega)+\bar{\varphi}_{cl}(\omega)A^2\Sigma^+_u\varphi_q(-\omega)\nonumber\\&+\bar{\psi}_{q}(\omega)(\Sigma^-_{ph}+B^2\Sigma^-_{u}-2B\Sigma^-_{ph-u})\psi_{cl}(-\omega)+\bar{\psi}_{q}(\omega)(A\Sigma^+_{ph-u}-AB\Sigma^+_{u})\varphi_{cl}(-\omega)\nonumber\\&+\bar{\varphi}_{q}(\omega)(A\Sigma^-_{ph-u}-AB\Sigma^-_{u})\psi_{cl}(-\omega)+\bar{\varphi}_{q}(\omega)A^2\Sigma^-_u\varphi_{cl}(-\omega)\nonumber\\&\bar{\psi}_{q}(\omega)2if_c(\mathrm{Im}(\Sigma_{ph}^+)+B^2\mathrm{Im}(\Sigma_{u}^+)-2B\mathrm{Im}(\Sigma_{ph-u}^+)\psi_{q}(-\omega))\nonumber\\&+\bar{\psi}_q(\omega)2if_c(A\mathrm{Im}(\Sigma_{ph-u}^+)-AB\mathrm{Im}(\Sigma_{u}^+))\varphi_q(-\omega)+\bar{\varphi}_q(\omega)2if_c(A\mathrm{Im}(\Sigma_{ph-u}^+)-AB\mathrm{Im}(\Sigma_{u}^+))\psi_q(-\omega)\nonumber\\&+\varphi_q(\omega)\mathrm{Im}(\Sigma_{u}^+)\varphi_q(-\omega)]\,,
\end{align}
\end{subequations}
where we define
\begin{subequations}\label{eq:B3}
\begin{align}
\Sigma_{ph}^{\pm}=&\mathcal{P}\int\frac{d\xi~\rho_r(\xi)\Gamma^2_{ph}(\xi)}{\xi-\omega}\pm i\pi\rho_r(\omega)\Gamma^2_{ph}(\omega)\\
\Sigma_{ph}^{\pm}=&\mathcal{P}\int\frac{d\xi~\rho_r(\xi)\Gamma^2_{u}(\xi)}{\xi-\omega}\pm i\pi\rho_r(\omega)\Gamma^2_{u}(\omega)\\
\Sigma_{ph-u}^{\pm}=&\mathcal{P}\int\frac{d\xi~\rho_r(\xi)\Gamma_{ph}\Gamma_{u}(\xi)}{\xi-\omega}\pm i\pi\rho_r(\omega)\Gamma_{ph}(\omega)\Gamma_{u}(\omega)\,,
\end{align}
\end{subequations}
and~$\mathcal{P}$~indicates Cauchy principle value. One can simpilify the equations by assuming the bath to be independent of frequency, that is to limit the calculation to Markovian baths; then the real part of all~$\Sigma$'s takes the zero value. Defining~$\gamma_a\equiv\sqrt{\pi\rho_r}\Gamma_{ph}$~and~$\gamma_u\equiv\sqrt{\pi\rho_r}\Gamma_{u}$, one finds the Eqs.~(\ref{eq:sDecay1})and~(\ref{eq:sDecay3}) of the main text.
\end{widetext}
\bibliographystyle{naturemag}
\bibliography{Sci,books,arXiv} 
\end{document}